\documentstyle[11pt]{article}
 
\newcommand{\beq}{\begin{equation}}
\newcommand{\eeq}{\end{equation}}
\newcommand{\bea}{\begin{eqnarray}}
\newcommand{\eea}{\end{eqnarray}}
\newcommand{\half}{\frac{1}{2}}
 
\begin{document}
\title{A Simple Model for Predicting Sprint Race
Times Accounting for Energy Loss on the Curve}
 
\author{J.\ R.\ Mureika\thanks{newt@sumatra.usc.edu} ~
\\
{\it Department of Computer Science} \\
{\it University of Southern California} \\
{\it Los Angeles, California 90089~~USA}}

\maketitle
\vskip .25 cm
 
\noindent
{\footnotesize
{\bf Abstract} \\
The mathematical model of J.\ Keller for predicting World Record race
times, based on a simple differential equation of motion, predicted quite well
the records of the day.
One of its shortcoming is that it neglects to account for a
sprinter's energy loss around a curve, a most important consideration
particularly in the 200m--400m.  An extension
to Keller's work is considered, modeling the aforementioned energy loss as a 
simple function of the centrifugal force acting on the runner around
the curve.  Theoretical World Record performances for indoor and outdoor
200m are discussed, and the use of the model at 300m is investigated.  
Some predictions are made for possible 200m outdoor and
indoor times as run by Canadian 100m WR holder Donovan Bailey, based
on his 100m final performance at the 1996 Olympic Games in Atlanta.
}

\section{Introduction}

In 1973, mathematician J.\ Keller \cite{kel} proposed a model for
predicting World Record (WR) race times based on a simple least--square
fit of the records of the day.  The fit was quite good, and provided
a simple tool for gauging possible optimual performances in  races, based 
on results from others.   Keller's model was limiting in the sense
that it could only ``in reality'' predict possible records of linear
races, with no consideration for those run on curves.  For distance
races (over 400m), this correction is negligible.  When the
race speeds are much higher, though, the curve contributions cannot
be left out. 

Recent WR performances in athletics have prompted various new studies
based on Keller's work.  The author of  \cite{tibs} introduces a more
realistic energy--loss model for sprinting, accounting for the sprinter's
actual velocity curve.  Still, though, the curve is not considered; this
is mentioned in \cite{tibs}, but no solution is offered.  The following
work will formulate a simple model to account for energy loss around
the curve, and predict possible WR performances accordingly, using data
obtained from a least--square fit of contemporary short sprint records.  
Both outdoor races, as well as indoor competitions, are discussed.
As a practical example, the 100m WR sprint race of Donovan Bailey (Canada)
is used as empirical data to further determine the validity of the
model for predicting 200m sprint times.  A brief discussion of indoor 300m
records is offered.  The possibility of using such a model as a training
tool for athletes and coaches in considered.

\section{The Keller Model}
 
        Although mathematical models for running were first introduced by
 A.\ V.\ Hill \cite{hill} in the mid--1920s, it was J.\ Keller who
formulated a model to predict possible WR performances \cite{kel},
based on the notion that the speed and energy loss of a human can be 
determined by certain key variables.  In its simplest form, the Keller
(or Hill--Keller) model is a solution to the simple equation of motion
 
\beq
\dot{v}(t) = f(t) - \tau^{-1} v(t)~,
\label{hk}
\eeq
Here, $f(t)$ is the force per
unit mass exerted by the runner, and $\tau$ is a decay 
constant which models internal (physiological) resistances felt by the runner.  
The differential equation (\ref{hk}) is solved subject to the
constraint $v(0) = 0$, and also bearing in mind that $f(t) \leq f$ 
({\it i.e.} the runner can only exert so much force).  The length of the
race $d$ can be calculated as

\beq
d = \int_0^T dt \, v(t)~,
\label{dist}
\eeq
and the time T to run the race can be obtained for a particular velocity
curve $v(t)$ over $d$.  An additional
constraint is that the power $f(t)\cdot v(t)$ must equal the rate of
internal energy supply (cellular oxygen replacement, anaerobic reactions,
{\it etc...}),
 
\beq
\frac{dE}{dt} = \sigma - f(t) v(t)~,
\label{hk2}
\eeq
with $\sigma$ a physiological term representing the body's energy
balance.  This is coupled with the initial condition $E(0) = E_0$,
as well as the non--negativity of $E(t)$ ($E(t) \geq 0$) \cite{kel}.
 
        By variational methods, it was determined \cite{kel} that
the optimal strategy for short sprints ($d < 291m = d_{crit}$) is for the runner
to go all--out for the duration of the race.  That is, $f(t) = f$.
Hence, $v(t)$ and $d$ can be calculated simply as
 
\bea
v(t) & = & f \tau (1- e^{-t/\tau})~, \nonumber \\
d & = & f \tau^2 \left(T/\tau + e^{-T/\tau} - 1\right)~.
\eea
For races of $d > d_{crit}$, the runner should chose a different optimization
strategy. The parameters determined by Keller are \cite{kel}
 
\bea
\tau & = & 0.892\;s \nonumber  \\
f & = & 12.2\;m/s^2 \nonumber  \\
\sigma & = & 9.83\; cal/(kg\;s) \nonumber  \\
E_0 & = & 575\; cal/kg  
\label{kelpar}
\eea
 
Keller \cite{kel} determined the optimal times (and hence WRs) for
short sprints, and found: 50m - 5.48s, 60m - 6.40s, 100m - 10.07s, 
200m - 19.25m, 400m - 43.27s.  Although  400m is beyond the short 
sprint category, this time is cited because of it's incredible approximation 
to the current record (43.29s, Harry ``Butch'' Reynolds, 1988).  Andre
Cason's (USA) 6.41s 60m WR is also surprisingly close.  
 
\section{Tibshirani's Extension}
 
        It is somewhat unrealistic to believe that a sprinter can actually 
apply a constant force for the duration of a race.  This being said, it seems 
logical to assume the force $f(t)$ decreases with time.  In the case of
\cite{tibs}, a linear decrease was chosen, $f(t) = f - ct$, where
$c > 0$.  In this case, the equations of motion become 
 
\bea
v(t)& =& k - c t \tau - k e^{-t/\tau}~, \\
D(t)& =& k t - \half c \tau t^2 + \tau k ( e^{-t/\tau} - 1 ) ~,
\label{tibseqs}
\eea
with $k = f\tau + \tau^2 c$.  
 
  More complex time dependences could equivalently be chosen
(for example, it might be more appealing to chose a time dependence of
the form $f(t) = f \exp(-\beta t)$),
but for the purposes of this study, the linar one will suffice.

\subsection{Accounting for reaction time}
 
The values in (\ref{kelpar}) were calculated without consideration of 
reaction time on the part of the athlete.  The IAAF sets the lowest possible
reaction time by a human being to be $t_{reac} = 0.100s$; any sprinter
who reacts faster than this is charged with a false start.  These times
generally do not drop below +0.130s, and in general register around
$+0.150$s (the average reaction time for the 100m and 200m finals at the
1996 Olympic Games was roughly +0.16s).   Granted, the ability to react
quickly is an important strategy, and obviously one which cannot really
be fit into a model.  At the 1996 Olympic Games, American sprinter Jon 
Drummond registered a reaction
time of +0.116s (100m, round 2, heat 2), and in the semi--final defending
champion Linford Christie (GBR) reacted in +0.124s \cite{ott}.
Such quick reactions tend to be more a result of anticipating the starter's
gun, though, rather than purely electrophysiological manifestations.

\section{Physical Meaning of the Parameters}
 
Although mathematical in origin, it is reasonable to hypothesize what might
be the physical interpretation of the parameters $(f, \tau, c)$.  Clearly,
$f$ is a measure of the raw acceleration capability of the sprinter,
while $f\tau$, having units of $m s^{-1}$, is a representation of velocity.
In fact, this is the maximum velocity which the sprinter is capable of
attaining (in the Keller model only; in the Tibshirani extension, the 
expression is slightly more complicated).  The variable $c$ must have units
of $f/t$, hence $m s^{-3}$.  Ideally, this is the time rate of change of
the runner's output, and can be thought of as a measure of muscular endurance.
The full implications of $\tau$ are unclear, but due to the nature of the
equation of motion, and keeping in mind the initial conjecture of Keller
that it be a function of internal resistances, one could hypothesize $\tau$
to be some type of measure of such elements as flexibility, leg turnover
rate, anaerobic responses, and so forth. 
 
While not necessarily representative of any {\it exact} physics quantity, 
these parameters may have some physical analogue.  The mechanics
of sprinting are far more complicated than the model suggests.  However, 
the mere fact that these models can predict race times with surprising 
accuracy indicates that perhaps they can be of some use in training.  One 
could imagine that a determination of the set $(f, \tau, c)$ for athletes 
can help to gear workouts toward specific development (power, endurance, 
and so forth).  Further investigation of the consistency of the model 
for various athletes might considered.

\section{200m races: Adjusting for the Curve}
\label{mymodel}
 
It is the opinion of this author that the way a sprinter handles the curve 
portion of a race, in particular a 200m, cannot be discounted.  Exactly
how this should be taken into consideration is unknown, as there are surely
various factors (both physical and physiological) which must be addressed.
The only physical difference between straight running and curve running
is obviously the effects of centrifugal forces on the sprinter.  One
can assume that a sprinter's racing spikes provide ample traction to 
stop outward translational motion, so this is not a concern.  To 
compensate for the rotational effects (torques), the sprinter leans into
the turn.  This is not constant during the race; greater speeds require
greater lean.  However, the degree of lean is limited by the maximum outward
angle of flexion of the ankle. Furthermore, one would think that maximum 
propulsive efficiency would not be generated at this extreme limit.  

So, a curve model is not a trivial one to construct.  However,
based on the physical considerations alone, let us assume that the effect
will manifest itself as a centrifugal term in the equation of motion.  
Since this is normal to the forward motion of the sprinter, we can rewrite 
(\ref{hk}) as
 
\beq
f(t)^2 = \left(\dot{v}(t) + {\tau}^{-1}v(t)\right)^2 + 
\lambda^2\frac{v(t)^4}{R^2},~
\label{mine1}
\eeq
The term $\lambda < 0$ has been added to account for the fact that a sprinter
does not feel the full centrifugal force resulting from his angular
velocity.   This seems to be the simplest choice, at least for a first approximation
to the correction.  Clearly, the Hill--Keller model is regained in the 
limit $R \rightarrow \infty$ (alternatively $\lambda \rightarrow 0$).

The radius of curvature $R$ can have two distinct sets of values, depending
on whether the competition is indoor or out,

\bea
R_{outdoor}& = &\left(\frac{100}{\pi} + 1.25 (p - 1) \right)m~, \nonumber \\
R_{indoor}& =& \left(\frac{50}{\pi} + 1.00 (p - 1) \right)m~, 
\label{radii}
\eea 
Here, $p$ is the lane number, and
the factors 1.25 (outdoor) and 1.00 (indoor)
have been chosen as suitable representations
of IAAF regulation lane widths, according to the following standards 
\cite{iaafsite}: 

\begin{itemize}
\item{{\underline{Outdoor:}} 400m in the inside lane, 
comprised of two 100m straights, and two 100m curves of fixed radius.  
Lane widths can range between 1.22 and 1.25 m, and are separated by lines 
of width 5 cm.}
\item{{\underline{Indoor:}} 200m in the inside lane (two 50m straights,
and two 50m curves).  The lanes (4 minimum, 6 maximum) should be between
0.90m to 1.10m in width, separated by a 5cm thick white line.  The
curve may be banked up to $18^o$, and should have a radius between
11m and 21m.  The radius need not be constant.}
\end{itemize}

        Solving Equation~(\ref{mine1}) for $\dot{v}(t)$, with $f(t) = f$,
one obtains
 
\beq
\dot{v}(t) = -\tau^{-1} v(t) + \sqrt{f^2 -
\lambda^2\frac{v(t)^4}{R^2}}~.
\label{mine2}
\eeq
Equivalently, for Tibshirani's more realistic model ($f(t) = f-ct$), 
Equation~(\ref{mine2}) becomes

\beq
\dot{v}(t) = -\tau^{-1} v(t) + \sqrt{(f-ct)^2-
\lambda^2\frac{v(t)^4}{R^2}}~.
\label{mine3}
\eeq
Because of a current lack of necessary empirical sprint data,
the value of $\lambda$ can only be estimated.
 
Differential equations of the form (\ref{mine2}), (\ref{mine3})
are not trivial to 
solve, as they yield no explicit solutions for $v(t)$.  However, such are 
easily solved by numerical methods.  This was performed on the MAPLE V 
Release 4 mathematical utility package, which uses a fourth-fifth order 
Runge--Kutta method.

The race distance $d$ traversed around the curve 
in time $T$ can be calculated analogously to 
Equation~(\ref{dist}),

\bea
d & = & d_c + d_s \\ \nonumber
& =& \int_0^{t_{1}} dt \, v_c(t) + \int_{t_{1}}^T dt \, v_s(t)~, 
\label{curvedist}
\eea
with $v_c(t)$ the solution to Equation~(\ref{mine3}), and $v_s(t)$ the
velocity as expressed in Equation~(\ref{tibseqs}), solved for the boundary
condition $v_c(t_1) = v_s(t_1)$. Here, $t_1$ is the time 
required to run the curved portion of the race (distance $d_c$), the integral 
form of which is evaluated numerically, based on the method of calculation 
stated for $v_c(t)$.

By using Keller's parameters (\ref{kelpar}), we can correct his original
prediction of 19.25s to account for the curve.   In fact, as an aside,
it should be mentioned that the record of 19.5s as indicated in
\cite{kel} is in fact the straight--track record of Tommie Smith, from
1966 \cite{iaafsite}.   With this in mind, we can apply the result of
(\ref{mine2}), coupled with (\ref{curvedist}), to obtain a curved--track
WR estimate.  For an assumed $\lambda^2 = 0.60$ (see section~\ref{db200}
for a discussion on choice of its value):

\bea
v_{100}& =& 10.66\,m/s~, \nonumber \\
t_{100}& =& 10.24\,s~,  \\
t_{200}& = & 19.46\,s~.\nonumber
\eea
The IAAF notes that times run on curves were estimated to be 0.3 to 0.4s slower
than straight runs \cite{iaafsite}.  These results would tend to agree with
this assertion.

\section{New Model Parameters for Modern World Records}
 
The parameters (\ref{kelpar}) are more than likely out of date, as they
were calculated by fitting records almost 25 years old \cite{kel}.
Also, these were fitted for a model which does not accurately model
the velocity curves of sprinters.  For example, a 100m runner's velocity
is not strictly increasing, but rather peaks between 40 and 60m.
Table~\ref{newwr} lists the sprint WRs as of March 1997, from 50m
to 400m \cite{iaafsite,ott}.  

New parameters $(f,\tau)$ and $(f,\tau,c)$ have been obtained by a 
least--square fit to
the four straight--track sprint WRs (50m, 55m, 60m, and 100m), and are
listed in (\ref{newkelpar},\ref{newpar}).  These reproduce the short
sprint times quite well (Table~\ref{newfit}).  Aside from
the 100m WR (where the reaction time is known, $t_{react} = +0.174s$
\cite{athann}), a (perhaps liberal) reaction time of $+0.16s$ has been 
assumed.  By using
the indoor races to calculate parameters, one is inherently removing
the possibility of wind--assisted times.  This has not been done in
the case of the 100m WR (where the wind--reading was $+0.7\;$m/s 
\cite{athann}), which may provide some source of 
error\footnote{\footnotesize Prictchard \cite{pritch1} 
offers a simple method of accounting for wind assistance and drag.  Making
use of his work, one finds that in fact Donovan Bailey's 9.84s WR corrects
to a 9.88s still--wind reading.  This is
surpassed by Frank Fredricks 9.86s run with a wind reading of $-0.4$m/s,
which adjusts to roughly 9.84s \cite{me2}.  So, if we account for 
wind contributions, a similar time is obtained anyway.}

\bea
f&=&10.230~m/s^2 \\ \nonumber
\tau & = &1.147~s 
\label{newkelpar}
\eea
with (lower, upper) asymptotic 95\% confidence levels of $f=(10.060,10.399)$,
$\tau=(1.124, 1.170)$, and
\bea
f & = & 9.596~m/s^2 \nonumber \\ 
\tau & = & 1.274~s \\
c & = & 0.058~m/s^3  \nonumber
\label{newpar}
\eea
with (lower, upper) asymptotic 95\% confidence levels of $f=(8.290, 10.901)$,
$\tau=(0.981,1.567)$, $c=( -0.065, 0.180)$.

In light of the discussions of Tibshirani's extension with relation to
observed velocity curves, the parameters (\ref{newkelpar}) are cited
only for comparison with older values (although predictions using 
(\ref{newkelpar}) are offered in Table~\ref{newkelpred}, as a
comparison to Keller's results).  Otherwise,
this work will use only the parameters of (\ref{newpar}).

\section{Predicting the 200m World Record}
\label{new200}

By a straight application of the model as described above, it is possible
to obtain predicted WR times for the 200m sprint.  In addition, it seems
logical to obtain predictions for indoor 200m races, as well, where the
dynamics of curve sprinting should be more apparent.  For outdoor performances,
$d_c = 100$m in (\ref{curvedist}), and $d_s = 100$m, and this is the same
for all lane choices $p=1 - 8$.  Recall that $d_c$ is not the
{\em curve--length} for all lanes, only the distance run on the curve.
For indoor races, the total distance is calculated by

\beq
d = d_{c1} + d_s + d_{c2} + d_s~,
\eeq
where $d_{c1,2}$ depend on the lane choice.  Since standard indoor tracks
are 200m in lane 1, it follows that $d_{c1} = d_{c2} = d_s = 50$m.  The
radius obviously increases for subsequent lanes, and using (\ref{radii}),
one obtains $d_{c1} = 40.58$m and $d_{c2} = 59.42$m.  The latter value is
the total length of the curved portion of lane 4, while the former is the
distance run after the stagger.

For all tables, unless otherwise indicated the times listed will be
raw ({\it i.e.} minus reaction time).  Only the final race times include
reaction, as indicated in the column headings.

\subsection{Outdoor 200m}

Calculations using various values of increasing $\lambda$ ($\lambda^2$)
are detailed in Table~\ref{200out} and Table~\ref{200in}.  For outdoor races 
(Table~\ref{200out}), a $\lambda^2$ range of 0.50-0.80 has been used.
Before the 1996 Olympic Games, the estimated times given would have been
considered almost unbelievable.  However, in light of the current 200m
WR (at the time of writing), the times are not so far fetched.  
The 19s--barrier is on the verge of being broken for $\lambda^2 = 0.50$,
while for higher $\lambda^2$, the current WR is approached.  It is interesting
to note that, for $\lambda^2 = 1.00$, the model predicts a time of
19.30s, quite close to Michael Johnson's 19.32s.  These predictions
are ideally for zero--wind readings, while the 19.32s was assisted with
a wind of $+0.4$m/s. It is quite possible that Johnson will
again lower his 200m WR mark this coming summer (1997), so we could
very well see times in the range predicted in Table~\ref{200out}.

  As a comparison to Keller's prediction of 19.25s 
\cite{kel}, which can be considered a straight--track 200m ($\lambda^2 = 0$),
this model yields $t_{200} = 18.54s + 0.16s = 18.70s$, with a split of
9.67s (which is just the prediction for the 100m WR).

\subsection{Indoor 200m}

Indoor tracks have much shorter radii of curvature than do outdoor
tracks. The centrifugal forces acting on a sprinter will be much higher for
large $v_c$, so it makes sense that the value of $\lambda$ assigned to 
subsequent calculations should be lower than for outdoor ones.  This is 
physically
realized by banked turns on indoor tracks, which are generally 2--4 feet
at maximum height.  How much lower a value of $\lambda$ one should choose
probably depends on the height of the particular bank, so again no
accurate estimate can be made.  Due to the $R^{1}$ force dependence, then
a $\lambda$ ($\lambda^2$) ratio in the range of 2:1 (4:1) might be expected
for an outdoor:indoor ratio
(under the assumption that the average maximal velocity about the curve
is the same).  Accurate measurements time and velocity measurements at
the end of each race segment (curves and straights) have been calculated,
and accurate measurement of these quantities 
can help determine validity of the model (see Table~\ref{tk200}).

 Frank Fredricks of Namibia broke the 20s barrier indoors in 1996 
(see Table~\ref{newwr}), setting a new 200m indoor WR of 19.96s.  This
can be used to estimate possible values of $\lambda$ that could be used.
Clearly, any value under $\lambda^2 = 0.60$ is quite reasonable,
and in fact the 19.51s prediction for $\lambda^2 = 0.40$ is attractive,
as it does not seem beyond the realm of possibility.   This does not
follow the 4:1 ratio outlined above, however there is no real reason
to believe that is should.  The only real stipulation is that indoor
values of $\lambda$ should be smaller than outdoor ones.

\subsection{Can the 19s barrier be broken?}

Suppose that a value of $\lambda^2 = 0.60$ holds for outdoor performances
(this assumption is based on results of Section~\ref{db200}).  The predicted
200m record is 19.08s, assuming a reaction of +0.16s (Table~\ref{200out}).
The minimum possible time allowed without a false start being called
would be 19.02s (this, of course, assumes no wind speed, for which the
predictions have been made; if there is a sufficient legal tail--wind, 
the mark would certainly fall).  How should this athlete train in 
order to break the 19s barrier?

A $0.4\%$ increase in the value of $f$ would give a raw time of 18.85s,
with a 100m split of 9.94s ($v_{100}=11.10\;$m/s).  Whereas, a larger
decrease of 9\% in $c$ (greater ``endurance'') would yield a raw time of 
18.83s, with a marginally slower split of $t_{100} = 9.95$s, but a 
slighlty faster $v_{100}=11.11\;$m/s.  This is an extreme case, but does show
how the model parameters might be useful to athletes and coaches as a 
training gauge. 

Various articles \cite{mcgill, future} have made attempts to
predict the future trends of WR performances, and the former states
that a sub--19.0s 200m could be realized by 2040 (although it also predicts
a 100m time of 9.49s to match).  The authors of \cite{future} are more
optimistic, predicting a WR of 18.97s being set as early as 2004.  While
their prediction of 19.52s for 1977 is off, it might  be retroactively
made consistent by Michael Johnson's 19.32s WR from the 1996 Olympic Games.
If the predicted times of Table~\ref{200out} are near accurate, and considering
the simple argument above, then the 2004 projection may not be far off the
mark.

\section{Is the 300m Now a Short Sprint?}
 
Keller determined that the maximum distance over which an athlete could
run using the strategy $f(t) = f$ was $d_{crit} = 291$m \cite{kel}.
Likewise, physiologists have suggested that a human cannot run at full
speed for longer that 30s (see \cite{pritch1} and references therein).   
While the latter study is just
over 10 years old, one wonders whether or not $d_{crit}$ has dropped.
Alternatively, if a sprinter can run a sub--30s 300m, would this entail
that the different strategy used for races longer than 291m no longer
applies?
 
As with the 200m, Table~\ref{300in} outlines possible 300m record
times, as run in lane 4 of a standard indoor track.  Since the actual
(if there is one) value of $\lambda^2$ is unknown, a range of $0.30$--$0.50$ 
is chosen in light of the 200m results.  In the case of lane 4, the race is 
made up of the segments
 
\bea
d & = & d_{c1} + d_s + d_{c2} + d_s + d_{c3} + d_s  \nonumber \\
 & = & 31.16m + 50m + 59.42m + 50m + 59.42m + 50m~.
\eea
 
The estimated time for the first two choices of  $\lambda^2$ are
under the 30s barrier by more than half a second, 
while $\lambda^2 = 0.5$ yields a value
of 29.72s (with reaction). Comparison to the current WR of 32.19s 
(Table~\ref{newwr}) give time differentials of approximately 2.49s--3.06s!
The 300m times may be a product of a decaying fit to the data.
However, the time differentials cited
appear far too large to be manifestations of statistical error alone,
which would suggest that there is an additional mechanism (perhaps 
physiological in origin) at work over this distance.  This approach would
suggest that the 300m is still not a sprint, by the definition of Keller
\cite{kel}.

\section{A Practical Application: Donovan Bailey}
\label{dbparams}

On Saturday, July 27th, 21:00 EST, Donovan Bailey (DB) of Canada 
crossed the 100m finish line in a new WR time of 9.84s (+0.7 m/s wind).  
Thanks to excellent documentation of data from this
race, it is possible to find an ``exact'' solution\footnote{\footnotesize It 
is emphasized that, while it is possible to obtain an exact set of
values for the parameters, these are not DB's parameters, since the
model does not account for wind assistance and drag.}
to the equations \ref{tibseqs}, and hence solve them for the parameters
$(f,\tau,c)$.  The relevant data for the race is \cite{athann},

\bea
v_{max}& =& 12.1 m/s~,  \\
d_{v_{max}} & = & 59.50 m~, \\
v_{100} & = & 11.5 m/s~,
\eea
Since the system
equations used are different than Keller's, the maximum velocity will not
be simply $v_{max} = f\tau$.  The maximum value of $v(t)$ is found to be

\beq
v_{max} = f\tau + c\tau^2 \ln \left\{ \frac{c}{f/\tau + c} \right\}~.
\eeq
wth $dv(t_{max})/dt = 0$.
The values $(f,\tau,c) = (7.96,1.72,0.156)$ are thus obtained.
These can be
compared with those obtained in \cite{tibs} by a least--square fit to 
the official splits listed in Table~\ref{100splits}:
$(f,\tau,c) = (6.41, 2.39, 0.20)$.  Note that the higher value of $f$ and
lower values of $\tau, c$ are likely a manifestation of solution method
and accounting for reaction time.

\subsection{Predicting DB's 200m times}
\label{db200}

Using the parameters obtained in Section~\ref{dbparams}, and the 
model framework established in Section~\ref{mymodel}, 200m times
will be obtained for DB as run on both indoor  and outdoor tracks.
Resulting split times are ``raw'' ({\it i.e.} without reaction time),
but the final time will be given both with and without reaction time
(roughly 0.15s, which is faster than
his 1996 Olympic 100m final reaction time 0f 0.174s).

Table~\ref{db200out} shows calculated times and velocities for DB running
in lane 4 ($p=4$) for varying values of $\lambda^2$.   Since the actual
value of this parameter is unknown, in order to determine its possible
value predicted times will be matched with DB's past 200m performances.
While no conclusive value of $\lambda^2$ could be determined from
Section~\ref{200in}, perhaps DB's performances can help shed light.
The IAAF lists \cite{iaafsite} his best 200m clocking as 20.76s, with a 
20.39s wind--assisted performance, in 1994.  Assuming that his time
will be lower in 1997 (but most likely not world--class, or sub--20s,
due to his training as a 100m specialist), it wil be assumed that
DB is currently capable of running roughly 20.20--20.30s.
This would tend to favor a value of $\lambda^2$ between 0.50 and 0.70.
Predicted indoor performances are listed in Table~\ref{db200in}.

For indoor 200m, Bailey's performance seems to greatly suffer for large
$\lambda$, which further supports the claim of smaller values for indoor
tracks.  A 200m clocking above 21s is hardly expected by a world class
sprinter!  In fact, even the 21s times ($\lambda^2 = 0.40, 0.50$) seem
somewhat slow for the 100m WR holder.  These could suggest that the
indoor $\lambda$ be quite low ($\lambda^2 < 0.4$).

Analogous to Table~\ref{tk200}, segment times and velocities for DB have
been calculated, and are listed in Table~\ref{db200v}.

\section{Discussion and General Conclusions}

This model is not intended to serve as gospel of how sprinters perform;
surely, it is crude at best.  However, it can be used as a simple tool to 
gauge what kind of records might be expected, based on present performances.  
Due to the ``loose'' statistical fit of the data from lack of points, the 
WRs of Section~\ref{new200} may be somewhat overestimated.  DB's predicted 
performances
of Section~\ref{db200} are probably more representative of the possible
range of $\lambda^2$ values  that one might realistically expect, if such
a model holds.  That is, if he is capable of running the 200m in the 
range of 20.15--20.40s, then if $\lambda^2$ is the same for all runners,
possible values lie between $\lambda^2 = 0.40 - 0.60$.  A value of
$\lambda^2 =1.00$, while closely reproducing the current 200m WR,
is definitely wrong from this observational point of view:
it would greatly underestimate Bailey's potential ($t > 20.60s$ would
hardly be expected by a WR holding sprinter, regardless of specialization).

	 The following points should be considered, though:

\begin{itemize}
\item{$\lambda^2$ is not the same for indoor and outdoor races;
indoor tracks would favor lower $\lambda$, so long as they are banked}
\item{$\lambda^2$ may not be the same for all athletes; 200m specialists
handle turns with greater ease than 100m specialists. It may be an
indivudual parameter, like $(f,\tau,c)$.}
\item{due to physiological considerations (different posture assumed or
muscles/joints used, {\it etc...}), it seems more likely that the
values of $\tau$ and/or $c$ may change around the curve}
\item{if the effect is purely physical, then the individual lane records
should be strictly decreasing from lane 1 to lane 8.  The recorded records
(Table~\ref{lanerec}) suggest that a minimal race time is achieved around lane
3 or 4, contributing to the physiological nature of curve running.}
\end{itemize}

The results of this paper are limited by the availability of relevant data,
unfortunately.  It would perhaps be of future interest to investigate the
physical nature of the parameters $(f, \tau, c, \lambda)$ through study 
of various
athletes.  By knowing the effects of their variability on predicted times,
models such as these could perhaps be used as a new training tool to 
gauge and direct the training of World Class athletes.

\vskip .25 cm
 
\noindent
{\bf Acknowledgements}
 
I thank R.\ Mureika and R.\ Turner (Applied Statistics Centre, Dept.\ of 
Mathematics and Statistics, University of New Brunswick) for assisting with 
the least--square fits of the current WRs used in this work.  I also thank 
R.\ Tibshirani for 
various insightful discussions, and D.\ Bailey for providing superb 
experimental data from which to work. \\

\pagebreak

\begin{table}
\begin{center}
{\begin{tabular}{|c c c l l l|}\hline
Event & $t (s)$ & $v_w$ (m/s)  & Athlete & Location &Date \\ \hline\hline
50m & 5.56 &i &Donovan Bailey (CAN) & Reno, NV & 9 Feb 1996 \\ 
55m & 5.99 &i, A & Obadele Thompson (BAR) &  Colorado Springs, CO & 22 Feb 1997 
\\
60m & 6.41 &i & Andre Cason (USA) & Madrid, ESP & 14 Feb 1992 \\
100m & 9.84 & +0.7 & Donovan Bailey (CAN) & Atlanta, GA & 27 Jul  1996 \\
200m & 19.96  &i & Frank Fredricks (NAM) & Li\'{e}vin, FR & 18 Feb 1996 \\
 & 19.32 & +0.4 & Michael Johnson (USA) & Atlanta, GA & 1 Aug 1996 \\
300m & 32.19 &i & Robson daSilva (BRA) & Karlsruhe & 24 Feb 1989 \\
400m & 44.63 & i & Michael Johnson (USA) & Atlanta, GA & 4 Mar 1995 \\
 & 43.29 &  & Harry Reynolds (USA) & Zurich & 17 Aug 1988 \\ \hline
\end{tabular}}
\end{center}
\caption{Men's Sprint World Records as of March 1997.  Wind speed of `i' 
indicates indoor performance; `A' indicates performance at altitude.}
\label{newwr}
\end{table}

\begin{table}
\begin{center}
{\begin{tabular}{|c|c|c|c|c|}\hline
Event & $t_{race}$ & $t_{raw}$ & $t_{fit}$ (Keller)  & $t_{fit}$ (Tibs.--Keller) \\ \hline
50m & 5.56 & 5.40 & 5.40 & 5.40 \\ \hline
55m & 5.99 & 5.83 & 5.83 & 5.83 \\ \hline
60m & 6.41 & 6.25 & 6.26 & 6.25 \\ \hline
100m & 9.84 & 9.67 & 9.67 & 9.67 \\ \hline
\end{tabular}}
\end{center}
\caption{Model predictions of Men's Sprint WRs;  $t_{raw} = t_{race} -
t_{reac}$, where $t_{reac} =$ 0.16s for all races except 100m (where it
has a known value of 0.17s).}
\label{newfit}
\end{table}

\begin{table}
\begin{center}
{\begin{tabular}{|c l c l l|}\hline
Lane & Athlete & $t_{200}$ &  Location & Date \\ \hline
1 & John Carlos               USA   & 20.12A  &   Mexico City &16 Oct 68 \\
 &  Daniel Effiong            NIG   & 20.15   &   Zurich     & 04 Aug 93 \\
2 &       Robson da Silva           BRA  &   20.00   & Barcelona  & 10 Sep 89 \\
3 &      Michael Johnson           USA  &   19.32   &  Atlanta    & 01 Aug 96 \\
4 &      Pietro Mennea             ITA  &   19.72A  &  Mexico City& 12 Sep 79 \\
  &      Michael Johnson           USA  &   19.79   &  Goteborg   & 11 Aug 95 \\
5 &      Michael Johnson           USA  &   19.66   &  Atlanta    & 23 Jun 96 \\
6 &      Joe DeLoach               USA  &   19.75   &  Seoul      & 28 Sep 88 \\
7 &      Carl Lewis                USA  &   19.80   & Los Angeles &08 Aug 84 \\
8 &      Michael Johnson           USA  &   19.79   & New Orleans &28 Jun 92 \\ \hline
\end{tabular}}
\end{center}
\caption{World records by lane for 200m (from \cite{200lnrec}).}
\label{lanerec}
\end{table}

\begin{table}
\begin{center}
{\begin{tabular}{|c||c c|c|c|}\hline
$\lambda^2$&$v_{100}$&$t_{100}$&$t_{200}$&$t_{200}+0.16$ \\ \hline
0.50&11.36 & 9.88 & 18.44 & 18.60  \\ \hline
0.60&11.29 & 9.92 & 18.49 & 18.65 \\ \hline
0.70&11.23 & 9.95 & 19.52 & 18.68\\ \hline
0.80&11.17 & 9.99 & 18.57 & 18.73\\ \hline
\end{tabular}}
\end{center}
\caption{Keller parameter ($f=10.230,\tau=1.147$) predicted outdoor 200m World 
Records for various values of $\lambda^2$, assuming race is run in lane 4.  
$v_{100}$ is the velocity for the given split.}
\label{newkelpred}
\end{table}

\begin{table}
\begin{center}
{\begin{tabular}{|c||c c|c|c|}\hline
$\lambda^2$&$v_{100}$&$t_{100}$&$t_{200}$&$t_{200}+0.16$ \\ \hline
0.50&11.14 & 9.92 & 18.86 & 19.02  \\ \hline
0.60&11.06 & 9.97 & 18.92 & 19.08 \\ \hline
0.70&10.98 & 10.02 & 18.98 & 19.14\\ \hline
0.80&10.91 & 10.06 & 19.03 & 19.19\\ \hline
\end{tabular}}
\end{center}
\caption{TK parameter ($f=9.596,\tau=1.274,c=0.058$) predicted outdoor 200m 
World Records for various values of $\lambda$, assuming race is run in lane 4.
$v_{100}$ is the velocity for the given split.}
\label{200out}
\end{table}

\begin{table}
\begin{center}
{\begin{tabular}{|c||c|c|c|c|c|}\hline
$\lambda^2$&$t_{50}$&$t_{100}$&$t_{150}$&$t_{200}$&$t_{200}+0.16$ \\ \hline
0.20&5.50 & 9.82 & 14.40 & 18.98 & 19.14 \\ \hline
0.30&5.55 & 9.88 & 14.56 & 19.17 & 19.33 \\ \hline
0.40&5.60 & 9.95 & 14.72 & 19.35 & 19.51 \\ \hline
0.50&5.64 & 10.01 & 14.86 & 19.52 & 19.68 \\ \hline
0.60&5.69 & 10.08 & 15.00 & 19.68 & 19.84 \\ \hline
\end{tabular}}
\end{center}
\caption{Predicted indoor 200m World Records for various values 
of $\lambda$, assuming race is run in lane 4.} 
\label{200in}
\end{table}

\begin{table}
\begin{center}
{\begin{tabular}{|c||c|c|c|c|c|c|c|}\hline
$\lambda^2$&$t_{50}$&$t_{100}$&$t_{150}$&$t_{200}$&$t_{250}$&$t_{300}$&$t_{300}+0.16$ \\ \hline
0.30&5.52 & 9.87 & 14.55 & 19.13 & 24.08 & 28.99 & 29.15 \\ \hline
0.40&5.55 & 9.93 & 14.69 & 19.29 & 24.33 & 29.27 & 29.43 \\ \hline
0.50&5.59 & 10.00 & 14.85 & 19.47 & 24.59 & 29.56 & 29.72 \\ \hline
\end{tabular}}
\end{center}
\caption{Predicted indoor 300m World Records, as run in lane 4.}
\label{300in}
\end{table}

\begin{table}
\begin{center}
{\begin{tabular}{|l||c c c c c c c c c c|}\hline
Split&10m&20&30&40&50&60&70&80&90&100 \\ \hline\hline
Speed&9.32&10.95&11.67&11.99&12.10&12.10&11.99&11.85&11.67&11.47 \\ \hline
Raw&1.89&2.90&3.79&4.64&5.47&6.29&7.12&7.96&8.81&9.67 \\ \hline
$+$reaction& 2.06&3.07&3.96&4.81&5.64&6.46&7.29&8.13&8.98&9.84 \\ \hline
Official&1.9&3.1&4.1&4.9&5.6&6.5&7.2&8.1&9.0&9.84 \\ \hline
\end{tabular}}
\end{center}
\caption{Predicted splits (s) and speed (m/s) compared with official for 
Bailey's 100m final in Atlanta.  Reaction time is rounded to $+$0.17s.}
\label{100splits}
\end{table}
 
\begin{table}
\begin{center}
{\begin{tabular}{|c||c c|c c| c| c| c|}\hline
$\lambda^2$&$t_{50}$&$v_{50}$&$t_{100}$&$v_{100}$&$t_{150}$&$t_{200}$&$t_{200}+0
.16$\\ 
\hline
0.25 & 5.53 & 11.74 & 9.89 & 11.03 & 14.56 & 19.81 & 19.97 \\ \hline
0.36 & 5.55 & 11.60 & 9.98 & 10.85 & 14.69 & 19.96 & 20.12 \\ \hline
0.50 & 5.59 & 11.43 & 10.09 & 10.65 & 14.84 & 20.13 & 20.29 \\ \hline
0.60 & 5.61 & 11.31 & 10.16&10.51&14.93&20.24&20.40 \\ \hline
0.70 & 5.63 & 11.20 & 10.24 & 10.39 & 15.09 & 20.43 & 20.59 \\ \hline
\end{tabular}}
\end{center}
\caption{Bailey's predicted outdoor 200m times, 
as run in lane 4.} 
\label{db200out}
\end{table}

\begin{table}
\begin{center}
{\begin{tabular}{|c||c|c|c|c|c|}\hline
$\lambda^2$&$t_{50}$&$t_{100}$&$t_{150}$&$t_{200}$&$t_{200}+0.16$\\ \hline
0.20& 5.62 & 9.91 & 14.88 & 20.32 & 20.48 \\ \hline
0.30& 5.68 & 10.01 & 15.17 & 20.71 & 20.87 \\ \hline
0.40& 5.75 & 10.13 & 15.43 & 21.05 & 21.21 \\ \hline 
0.50& 5.81 & 10.22 & 15.67 & 21.37 & 21.53 \\ \hline 
0.60& 5.88 & 10.32 & 15.91 & 21.68 & 21.84 \\ \hline
0.70& 5.94 & 10.42 & 16.13 & 21.97 & 22.13 \\ \hline
0.80& 5.99 & 10.50 & 16.33 & 22.23 & 22.39 \\ \hline
\end{tabular}}
\end{center}
\caption{Bailey's predicted indoor 200m times, as run in lane 4.}
\label{db200in}
\end{table}

\begin{table}
\begin{center}
{\begin{tabular}{|c||c|c||c|c||c|c|}\hline
$\lambda^2$&$t_{c1}$&$v$&$t_{s1}$&$v$&$t_{c2}$&$v$ \\ \hline
0.20 & 4.67 & 11.16 & 8.99 & 11.63 & 14.40 & 10.70 \\ \hline
0.30 & 4.71 & 10.95 & 9.05 & 11.62 & 14.56 & 10.47 \\ \hline
0.40 & 4.75 & 10.76 & 9.11 & 11.61 & 14.72 & 10.27 \\ \hline
0.50 & 4.78 & 10.59 & 9.16 & 11.60 & 14.86 & 10.09 \\ \hline
0.60 & 4.82 & 10.43 & 9.22 & 11.59 & 15.00 & 9.92 \\ \hline
\end{tabular}}
\end{center}
\caption{TK parameter times and velocities for curve ($c1=40.58$m, $c2=59.42$m),
and straight ($s1 = 50$m) race segments for indoor 200m.}
\label{tk200}
\end{table}

\begin{table}
\begin{center}
{\begin{tabular}{|c||c|c||c|c||c|c|}\hline
$\lambda^2$&$t_{c1}$&$v$&$t_{s1}$&$v$&$t_{c2}$&$v$ \\ \hline
0.20 & 4.79 & 11.20 & 9.07 & 11.58 & 14.88 & 9.40 \\ \hline
0.30 & 4.84 & 10.90 & 9.17 & 11.53 & 15.17 & 9.06 \\ \hline
0.40 & 4.89 & 10.62 & 9.26 & 11.49 & 15.43 & 8.80 \\ \hline
0.50 & 4.94 & 10.38 & 9.34 & 11.46 & 15.67 & 8.56 \\ \hline
0.60 & 4.99 & 10.15 & 9.43 & 11.42 & 15.91 & 8.35 \\ \hline
\end{tabular}}
\end{center}
\caption{Bailey parameter times and velocities for curve ($c1=40.58$m, $c2=59.42$m),
and straight ($s1 = 50$m) race segments for indoor 200m.}
\label{db200v}
\end{table}

\end{document}